\documentclass{emulateapj}
\shorttitle{BAO on 21cm emission}
\shortauthors{Mao \& Wu}

\begin{document}

\title{Signatures of the Baryon Acoustic Oscillations on 21cm Emission 
Background}

\author{Xiao-Chun Mao\altaffilmark{1,2} and Xiang-Ping Wu\altaffilmark{1}}
\altaffiltext{1}{National Astronomical Observatories, Chinese Academy of
Sciences, Beijing 100012, China}
\altaffiltext{2}{Graduate School of Chinese Academy of Sciences, 
Beijing 100049, China}

\begin{abstract}
The baryon acoustic oscillations (BAO) prior to recombination
should be imprinted onto the 21cm emission background 
from the epoch of reionization through the underlying density 
perturbations. Using an analytical approach for both matter
power spectrum (CDM+baryons) and reionization process, we 
demonstrate the BAO induced signatures on the power spectrum of
21cm emission fluctuations. Future low-frequency radio 
telescopes such as LOFAR and MWA should be able to
detect these weak BAO wiggles with an integration time of $\sim1$ 
year. A combination of the BAO measurements at different
redshifts $z\approx1000$ (CMB), $z\approx10$ (epoch of
reionization) and $z\approx0$ (clustering of galaxies) may allow
one to set more robust constraints on the determinations of
cosmological parameters including dark energy and its equation
of state.  
\end{abstract}


\section{Introduction}

Prior to recombination, free electrons coupled the baryons and photons 
tightly through Compton scattering, and these three species moved 
together as a single fluid. In this relativistic plasma, the primordial 
small-scale perturbations propagated as sound waves, resulting in the 
pressure-induce oscillations. The neutral gas can still retain some 
memory of such acoustic oscillations even after recombination, 
manifesting themselves in the last scattering surface seen as
the harmonic series of maxima and minima on the cosmic microwave 
background (CMB) at redshift $z\approx1000$. The longest wavelength 
of BAO with $\lambda\approx100$ Mpc imprinted on the large-scale 
structures is still visible in the local universe 
through the survey of 3D galaxy distributions \citep{Eisenstein05,Cole05}.

Because BAO can be served as an ideal cosmic ruler for many cosmological 
applications especially for the probe of dark energy 
\citep{Hu96,Barkana05,Wyithe07}, it is of great interest to explore
how BAO evolve with cosmic time, and in particular how and when the 
peaks of BAO at smaller scales are washed out by emergence of larger 
structures. Indeed, in addition to the detections of BAO signatures 
in CMB at $z\approx1000$ and galaxy spatial distributions at lower redshifts,
one may be able to extract valuable information on BAO at redshifts around  
$z\approx10$ from the study of 21 cm absorption/emission 
generated in the dark ages and epoch of reionization. 
This will complement our knowledge of structure formation
at this important phase when galactic dark halos started to develop.   
The existing modes and positions of BAO at the epoch of reionization 
would make a sensitive diagnosis of nonlinear structures evolved  
by that time. Most importantly, the statistical uncertainties in
the determination of cosmological parameters including dark
energy and its equation of state can be significantly reduced when
more independent measurements of BAO through the epoch of reionization 
are incorporated with the BAO features already detected in CMB and 
large-scale structures of the local universe. Note that the 21 cm 
absorption/emission observations provide a tomographic imaging of
the universe at the epoch of reionization and dark ages, yielding
many independent constraints on the theory of cosmology.

The goal of this letter is to demonstrate the signatures 
of BAO on the redshifted 21 cm fluctuations and discuss the 
feasibility of detections with existing and planned low  
frequency telescopes such as 21CMA, LOFAR and MWA. We focus on 
the 21cm emission of neutral hydrogen during the process of 
reionization instead of the 21 cm absorption at $z>20$ \citep{Barkana05}. 
For the latter, the detection of BAO signatures turns to be much more 
difficult because of their longer wavelength and limitation of angular 
resolutions with exiting and even future radio telescopes. 
A sophisticated treatment of the problem requires the detailed knowledge of  
the history of reionization and radiative transfer of ionizing photons 
through gas density field, which may be achievable by numerical
simulations \citep[e.g.][]{Furlanetto04}. 
Here we would rather employ an analytic approach to
simplify the problem and highlight the essentials of physical 
process at the epoch of reionization.   Throughout the 
paper we adopt a concordance cosmology of 
$\Omega_0=0.265$, $\Omega_\Lambda=0.735$, $\Omega_b=0.044$, 
$h=0.71$, $n_s=1$ and $\sigma_8=0.772$, as revealed by the WMAP 
three-year observations \citep{Spergel07}.

\section{Matter Power Spectra}

We begin with the linear matter power spectrum 
$P_{\rm lin}(z,k)\propto D^2(z)k^{n_s}T^2(k)$,
which describes how the initial matter power spectrum $k^{n_s}$ is
modulated by the transfer function $T(k)$ and growth factor 
$D(z)$. The baryon content, and thereby BAO, is incorporated in 
$T(k)$ which can be approximately separated into the cold dark matter
(CDM) and baryon components:
$T(k)=(\Omega_b/\Omega_0)T_b(k)+(\Omega_c/\Omega_0)T_c(k)$,
where $\Omega_c$ is the CDM density parameter relative to the critical 
density at present, $\Omega_c=\Omega_0-\Omega_b$.
We adopt the asymptotic solutions to both $T_b(k)$ and $T_c(k)$
near the sound horizon given by \citet{Eisenstein98}, 
in which the suppression effect of baryons on scales
below the sound horizon is included. Nonetheless, the linear matter 
power spectrum becomes inaccurate at smaller scales and later 
cosmic time. We employ a halo model to evaluate the nonlinear
power spectrum \citep[see][]{Cooray02}, which accounts for 
contributions from the single halo term $P_{\rm 1h}$ plus the clustering 
term  $P_{\rm 2h}$. We take the Press-Schechter formalism for the
mass function of dark halos and the Navarro-Frenk-White \citep{Navarro96} 
profile for the matter distribution inside each halo.    
Employment of the nonlinear matter power spectrum instead of 
the linear one is crucial in the sense that BAO can be 
erased entirely on scales of nonlinear structures, although this
may not be a serious problem at the epoch of reionization.

\begin{figure}
\begin{center}
\includegraphics[angle=270, scale=0.35]{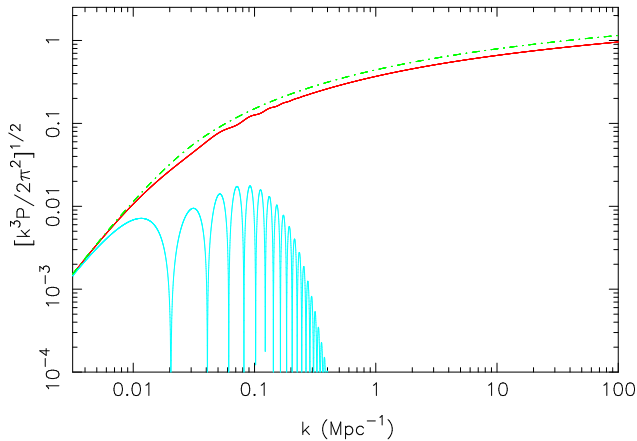}
\caption{Linear matter power spectra at $z=6$ for three different matter 
contents: pure CDM (dot-dashed line), pure baryons (grey) and 
mixed CDM+baryons (solid line). The BAO features can be seen at 
$k\sim0.1$ Mpc$^{-1}$ in the mixed matter model.}
\end{center}
\end{figure} 

In Fig.1, we demonstrate the linear matter power spectra $P_{\rm lin}$ 
at redshift $z=6$ for three matter contents: pure CDM, pure baryons 
and mixed CDM+baryons. Inclusion of baryons in the pure CDM model 
gives rise to the weak yet visible wiggles in the matter power 
spectrum at wavenumber $k\sim0.1$ Mpc$^{-1}$.  
Next, we calculate the nonlinear matter power spectrum instead of $P_{\rm lin}$
but leave the baryon content unchanged, and the result for
$z=6$ is shown in Fig.2.  
The nonlinear matter power spectrum at large scales $k<1$ Mpc$^{-1}$
remains roughly the same as $P_{\rm lin}$, in which the weak BAO are 
clearly presented.  The prominent nonlinear structures dominate
the matter power spectrum only at short wavelengths of $k>1$ Mpc$^{-1}$.
It turns out that by the end of cosmic reionization at $z\approx6$, 
BAO are still unaffected by the formation of nonlinear structures.
This arises because the gravitationally bound systems such as 
dark halos and their associated large-scale structures at $z=6$  
have sizes much smaller than the typical scales ($\sim100$ Mpc) of BAO.  
In other words, many of the interesting modes of BAO should 
leave their imprints on the matter power spectrum before $z=6$.

\begin{figure}
\begin{center}
\includegraphics[angle=270, scale=0.35]{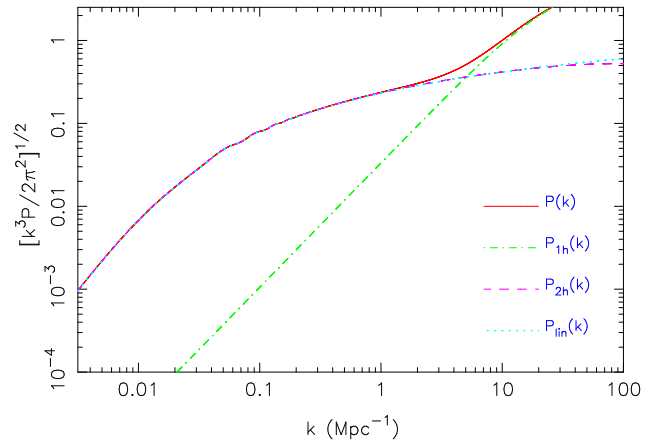}
\caption{The matter power spectra at $z=6$ predicted by linear density
perturbation (dotted line) and halo model (solid line). For
the latter, contributions of 1-halo and 2-halo terms are 
also displayed.  Linear theory breaks down at small scales
with wavenumber beyond $k>1$ Mpc$^{-1}$, and BAO are unaffected by
nonlinear structures by $z=6$.  }
\end{center}
\end{figure}

\section{21cm Power Spectra}

BAO signatures enter into the redshifted 21cm emission background
from the epoch of reionization through the
underlying matter density fluctuations ($\delta$). If we restrict ourselves
to the 21cm emission generated from the neutral hydrogen in the 
surroundings of the ionized bubbles of first-generation 
luminous objects, the surface brightness of the emission can
be evaluated through \citep[cf.,][]{Zaldarriaga04}
\begin{eqnarray}
\delta{T_b}&\approx& T_0(1+\delta)(1-x) \nonumber\\
T_0&=&16\,{\rm mK}\,h^{-1} \Bigg(\frac{\Omega_{\rm b}h^2}{0.02}\Bigg)
    \Bigg(\frac{1+z}{10}\frac{0.3}{\Omega_{\rm M}}\Bigg)^{1/2},
\end{eqnarray}
where $x=x_e(1+\delta_{x})$ is the ionization fraction, 
$x_e$ is the average ionization fraction and $\delta_{x}$ is the 
perturbation in the ionization fraction across the sky, for which 
we will take the reionization model of \citet{Santos05}.
The corresponding power spectrum of the 21cm emission can be 
written as 
\begin{eqnarray}
P_{21}^{3D}(z,k)&=&T_0^2\Big[(1-x_e)^2P_{\delta\delta}(z,k)+x_e^2P_{\delta_x
\delta_x}(z,k)\Big]\nonumber\\
&&-\Big[2x_e(1-x_e)P_{\delta\delta_x}(z,k)\Big].
\end{eqnarray}
The three terms in the right-hand side represent the contributions of 
the matter power 
spectrum, the power from the perturbations in the ionization fraction, 
and the cross-correlation power, respectively.  For the latter two, 
we use the model of \citet{Santos03} to proceed our numerical
computation.  The angular power spectrum  $C_\ell(\nu)$ of the redshifted 
21cm brightness sky is derived under the flat-sky approximation.

\begin{figure}
\begin{center}
\includegraphics[angle=270, scale=0.35]{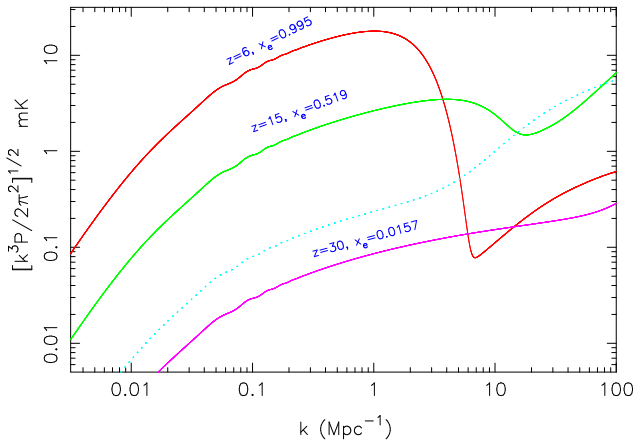}
\caption{Power spectra of 21cm fluctuations at different redshifts from
$z=30$ to $z=6$. The ionization fraction $x_e$ is also showed at
each redshift. BAO wiggles occur at $k\sim0.1$ Mpc$^{-1}$ and 
nonlinear structures dominate at $k>1$ Mpc$^{-1}$.  The matter
power spectrum at $z=6$ is also plotted for comparison (dotted line).} 
\end{center}
\end{figure} 

\begin{figure}
\begin{center}
\includegraphics[angle=270, scale=0.35]{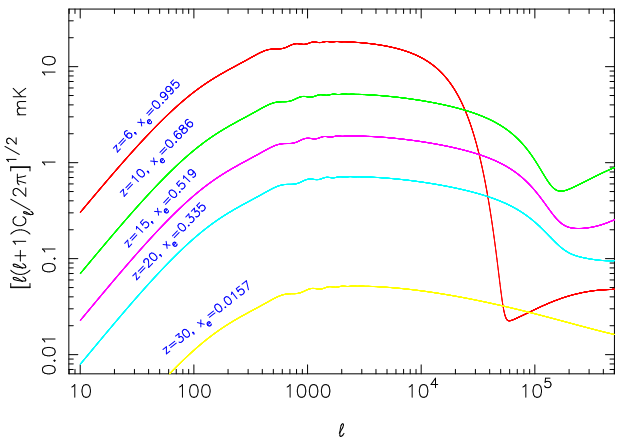}
\caption{Angular power spectra of 21cm fluctuations at different redshifts 
from $z=30$ to $z=6$. BAO wiggles are presented at $\ell\approx500-3000$,
and their positions tend towards large $\ell$ with the increase of redshift. }
\end{center}
\end{figure} 

The theoretically predicted 3D and 2D power spectra of the redshifted 21cm 
emission fluctuations are shown in Fig.3 and Fig.4, respectively. 
While the amplitudes of the 21cm power spectra themselves are relatively
low, with a maximum value of $\sim10$ mK, the BAO induced wiggles
are clearly presented. It appears that  an angular resolution of
$\sim10$ arcminute is needed in order to identify the BAO features on the
2D power spectra, apart from the requirement of high sensitivity. 
Moreover, the positions of BAO tend towards large $\ell$ with the
increase of redshift, indicating that radio arrays with baseline
of $\sim1$ km will be needed to reveal these structures.
We have also calculated the 1D power spectrum of the 21cm emission along
the line of sight but found that the BAO features are completely washed 
out due to projection effect, in agreement with the result of \citet{Wang06a}.
Finally, it is pointed out that our analytical model does not take
the size distribution of ionized bubbles into account. 
The typical sizes of the ionized bubbles near $z\approx6$ can reach 
$\sim20$ Mpc, which is already comparable to the BAO wiggles at
large $k$ or $\ell$. Whether or not the ionized bubbles produce 
oscillations on the same scales as BAO should be investigated 
in future study.

\section{Detectability}

Detection of the BAO signatures on the redshifted 21cm fluctuations 
indeed poses a technique challenge for existing and planned low-frequency
radio telescopes. We demonstrate the observability 
using 21 CentiMeter Array (21CMA, {\it cosmo.bao.ac.cn}), 
Low Frequency Radio Array (LOFAR, {\it www.lofar.org}) 
and Mileura Widefield Array (MWA, {\it www.haystack.mit.edu/arrays/MWA}), 
and only work with the angular power spectrum $C_{\ell}$ at a fixed frequency. 
Supposing that strong foreground contamination can be entirely removed from 
the low frequency sky through either the two-point correlation technique 
in frequency domain \citep[e.g.][]{Zaldarriaga04} or the pixel-by-pixel  
algorithm \citep{Wang06b}, we can estimate the variance in $C_{\ell}$ through
$\bigtriangleup C_\ell= [2/(2\ell+1)f_{\rm sky}]^{1/2} (C_\ell+N_{\ell})$,
in which $f_{\rm sky}=\pi \theta_{\rm deg}^2/129600$ 
accounts for the sky coverage, and 
$N_{\ell}=(wf_{\rm sky})^{-1}e^{\theta_b^2\ell(\ell+1)}$ is the noise power 
spectrum if we adopt a Gaussian function with width $\theta_b$ for the 
experimental beam and use $w^{-1}=4\pi\sigma_{\rm pix}^2/N_{\rm pix}$ to 
denote the contribution of the  white noise with $\sigma_{\rm pix}$ 
and $N_{\rm pix}$ being the pixel noise and total number 
of pixels, respectively. In radio interferometric measurement, the 
pixel noise can be represented in terms of brightness temperature as 
$\sigma_{\rm pix}=T_{\rm sys}/\eta\sqrt{2N\Delta\nu t}$, where 
$T_{\rm sys}$ is the system temperature, $\eta$ is the efficiency factor
of telescope, $N$ is the total number of independent baselines, 
$\Delta\nu$ is the bandwidth, and $t$ is the observing time.

To proceed further, for 21CMA we take a system temperature of 
$T_{\rm sys}=250$ K and an efficiency of $\eta=0.64$. 
The total dishes of 21CMA are $N_{\rm dish}=81$ and the longest baseline
is 6 km, which gives rise to an angular resolution of  $\theta_b\approx1$ 
arcmin. We use a conservative value of  $\theta_b=2$ 
arcmin in the present estimate. The sky coverage is, nevertheless, 
very small: $f_{\rm sky}=10^{-3}$.   
For LOFAR (compact core), the corresponding parameters are chosen to be: 
$T_{\rm sys}=100$ K, $\eta=0.64$, $N_{\rm dish}=32$,
$\theta_b=3$  arcmin, and $f_{\rm sky}\approx0.1$. We utilize the following
parameters for MWA: $T_{\rm sys}=200$ K, $\eta=0.64$, 
$N_{\rm dish}=500$, $\theta_b=5$  arcmin, and $f_{\rm sky}\approx0.4$.
The errors $\Delta C_{\ell}$ in the measurement of the 21cm power spectra 
with 21CMA, LOFAR (core) and MWA are displayed in Fig.5 for 
$z=10$ and $z=20$, respectively. 
While it is promising for all the three experiments to detect 
the reionization signals in the angular power spectra of 21cm
fluctuations over a wide range of redshifts beyond $z=6$ and angular
scales from  $\ell\sim10^2$ to $\sim10^4$, after an integration time 
of $\sim1$ year, a significant detection of
the BAO wiggles on the 21cm angular power spectrum turns to be still
difficult especially at higher redshifts (or lower frequencies) 
due to both the weak signals of BAO themselves and the limitation of 
angular resolutions of current radio telescopes. The strategy is that 
the maximum variations of BAO (e.g. the power difference between 
acoustic peaks and adjacent troughs) 
should exceed the errors at the corresponding modes. To optimize 
the detection, one should choose to work at higher frequencies
near 200MHz, though the total 21cm signals may become weaker because the
reionization process almost completed by $z=6$. To be specific, 
the relatively smaller sky coverage and higher system temperature of 
21CMA prevent a significant detection of the BAO wiggles 
on the angular power spectra of 21cm fluctuations unless a longer
integration time of $\sim10$ years is allowed.   
In contrast, both LOFAR and MWA may be able to capture 
the BAO signals within $\sim1$ year. In particular, MWA can even 
trace all the BAO wiggles out to $z\approx30$ and on very small angular 
scales of a few arcminutes because of its numerous independent baselines.
We anticipate that a similar result can be reached when all the 77 stations 
in the planned LOFAR start to operate.

\begin{figure}
\begin{center}
\includegraphics[angle=270, scale=0.35]{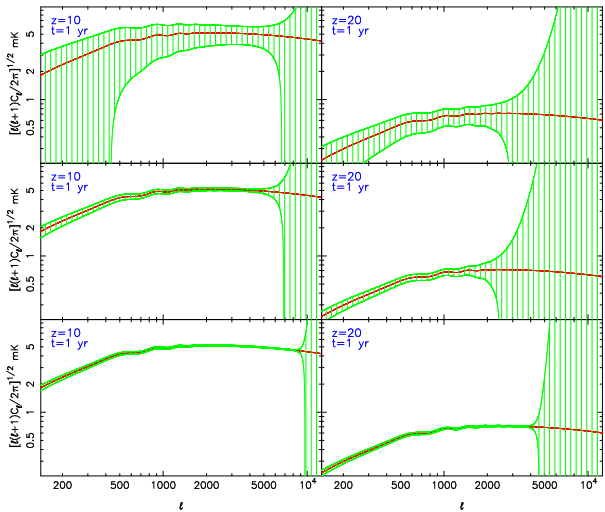}
\caption{The expected angular power spectra and measurement errors 
with 21CMA\,(top), LOFAR(core)\,(middle) and MWA\,(bottom) for  
a bandwidth of $\Delta\nu=0.1$ MHz and an integration time of 1 year.}
\end{center}
\end{figure}

\section{Conclusions}

The BAO should be imprinted onto the 21cm emission background 
from the epoch of reionization through the underlying density 
perturbations. Detection of the signals will provide valuable
information about the formation and evolution of cosmic structures
at higher redshifts beyond $z\approx6$. It also furnishes 
a standard ruler to the probe of topology and geometry of 
the universe including dark energy and its equation of state.   
In particular, many of the BAO modes were not erased by the 
formation of large-scale structures by $z=6$, and we may be
able to see the BAO wiggles at smaller scales.  This will
complement our knowledge of the BAO at intermediate redshifts
in addition to the detections of BAO signatures in CMB at
$z\approx1000$ and in large-scale distribution of galaxies at $z\approx0$.
A combination of these BAO measurements at different redshifts
will allow us to set more robust constraints on the determinations
of cosmological parameters.

We have used an analytic approach based on halo model 
for distribution and evolution of dark matter, in which 
baryons trace essentially dark matter but the standard CDM
power spectrum is modified by the presence of baryons. 
We have then calculated the 21cm emission power spectrum 
from neutral hydrogen surrounding the ionized bubble of each halo, 
following a simple model of ionization history. 
Our results show that BAO are indeed presented at the
power spectra of the redshifted 21cm emission from the epoch 
of reionization, and are almost unaffected by 
the presence of nonlinear structures beyond $z>6$. This indicates
that one should be able to see many of the BAO modes at 
the 21cm power spectrum.

We have worked with the angular power spectra of the 21cm emission from
the epoch of reionization for a fixed frequency. The BAO
signatures are clearly seen at $\ell\approx500$-$3000$ 
through the entire history of reionization. However, detections of these
wiggles with existing and planned radio interferometric arrays 
such as 21CMA, LOFAR and MWA does pose a technique challenge. 
The primary difficulty, apart from the extremely faint signals 
of 21cm emission themselves from the epoch of reionization and 
the strong foreground at low frequency, arises from the high
system noise and the limitation of angular resolutions. 
Yet, within an integration time of about 1 year, it seems that 
both LOFAR and MWA are capable of capturing the BAO signatures  
at $\ell\sim1000$, provided that the foreground contamination 
can be successfully removed.

\acknowledgments{We thank an anonymous referee for valuable
comments and suggestions. This work was supported by 
the Chinese Academy of Sciences under grant KJCX2-YW-T02}

\clearpage

\end{document}